\documentclass[preprint,12pt]{elsarticle}
\usepackage{graphicx}
\usepackage{amssymb}
\usepackage{lineno}
\usepackage[english]{babel}
\usepackage[utf8]{inputenc}
\usepackage{amsmath}
\usepackage[nottoc]{tocbibind}
\usepackage{float}
\usepackage[font={footnotesize,sf}]{caption}

\journal{Extreme Mechanics Letter}

\begin{document}

\begin{frontmatter}

\title{Asymmetric Multi-Stability \\ from Relaxing the Rigid-Folding Conditions \\ in a Stacked Miura-ori Cellular Solid}

\author{Jiayue Tao \corref{cor1}}
\cortext[cor1]{Corresponding Author: \texttt{jiayuet@clemson.edu}}
\author{and Suyi Li}

\address{Department of Mechanical Engineering, Clemson University}

\begin{abstract}

Traditionally, origami has been categorized into two groups according to their kinematics design: rigid and non-rigid origami.  However, such categorization can be superficial, and rigid origami can obtain new mechanical properties by intentionally relaxing the rigid-folding kinematics.  Based on numerical simulations using the bar-hinge approach and experiments, this study examines the multi-stability of a stacked Miura-origami cellular structure with different levels of facet compliance.  The simulation and experiment results show that a unit cell in such cellular solid exhibits only two stable states if it follows the rigid origami kinematics; however, two more stable states are reachable if the origami facets become sufficiently compliant.  Moreover, the switch between two certain stable states shows an asymmetric energy barrier, meaning that the unit cell follows fundamentally different deformation paths when it extends from one state to another compared to the opposite compression switch.  As a result, the reaction force required for extending this unit cell between these two states can be higher than the compression switch.  Such asymmetric multi-stability can be fine-tuned by tailoring the underlying origami design, and it can be extended into cellular solids with carefully placed voids.   By showing the benefits of exploiting facet compliance, this study could foster multi-functional structures and material systems that traditional rigid origami cannot create.

\end{abstract}

\begin{keyword}
Non-rigid Origami \sep Multi-stability \sep Asymmetric Energy Barrier
\end{keyword}

\end{frontmatter}

\section{Introduction}

Over the past decade, we have witnessed the rapid emergence of multi-stable structures and material systems in many engineering disciplines.  These structures and materials possess more than one stable equilibrium (or stable state), and they can settle in any state without external aids.  Moreover, internal or external actuation can rapidly ``snap'' these structures and materials between different states through elastic instability \cite{Iniguez-Rabago2019}.   These unique nonlinear behaviors can be exploited to achieve a wide variety of adaptive functions like wave propagation control \cite{Nadkarni2016, Raney2016a}, energy harvesting \cite{Daqaq2014, Harne2013, Pellegrini2013}, shape morphing \cite{Daynes2013a, Lachenal2013, Sun2016b}, mechanical property adaptation \cite{Florijn2014, Kidambi2016, Wu2016a, Harne2016, Grima2013}, impact energy absorption \cite{Shan2015, Frenzel2016}, robust sensing \cite{Harne2014a, Harne2015}, and robotic tasks \cite{Kim2014a, Chen2018a, Bhovad2019a, Preston2019}. 

The mechanisms for achieving multi-stability are pretty diverse, including linkage-based structures, buckled beams, and pre-stressed shells \cite{Sengupta2018a,Lele2019a}.  Among them, origami --- the ancient craftsman art of paper folding --- has received much attention recently due to its conceptual simplicity, fabrication scalability, and design versatility \cite{Peraza-Hernandez2014, Johnson2017, Rus2018}.  Traditionally, origami has been divided into two categories according to the folding kinematics (aka. rigid and non-rigid origami), and both categories can show multi-stability based on different principles.   Folding of the rigid origami, by kinematics design, only requires bending along its creases without incurring any facet deformations.  Therefore, this kind of origami is essentially a three-dimensional linkage mechanism consisting of rigid facets connected via hinge-like creases (hence the name rigid or rigid-foldable origami).  As a result, multi-stability can arise via the combination of nonlinear folding kinematics and purposefully assigned spring hinge stiffness on the creases.  Examples of multi-stable rigid origami includes Miura-ori \cite{Waitukaitis2015}, TMP bellow \cite{Yasuda2015a}, leaf-out \cite{Yasuda2016}, and water-bomb base \cite{Sadeghi2020}.  On the other hand, folding of the non-rigid origami requires facet deformation, which can be used to achieve multi-stability.   The facets of non-rigid and multi-stable origami are relatively stress-free \emph{at} a stable state but undergo significant deformation when folding \emph{between} different states.  Examples of multi-stable non-rigid origami include Kresling \cite{Kaufmann2021a}, square twist \cite{Ma2021}, and the star pattern \cite{Kamrava2019, Melancon2021}.

However, the distinctions between rigid and non-rigid origami can be superficial.  Non-rigid origami can be approximated by an equivalent rigid-foldable pattern using ``virtual''  crease lines \cite{Silverberg2015,Pagano2017}, and rigid origami can exhibit non-rigid folding behaviors (aka. facet deformations) due to the inevitable material compliance and fabrication imperfections in practical applications \cite{Liu2018}.  This study focuses on the latter case.  Facet compliance in rigid origami can be undesirable because it complicates the overall deformation characteristics and reduces origami's load-bearing capacity.  However, facet compliance allows the rigid origami to access otherwise unachievable folding configurations.  As a result, new mechanical properties can arise. 

Therefore, we aim to examine how rigid origami can receive new multi-stability characteristics and functionality via intentionally relaxing the rigid-folding condition and exploiting facet compliance.  To this end, we use a variation of stacked Miura-ori cellular structure as the testbed.   Stacked Miura-ori, as its name implies, is a space-filling cellular structure consisting of different Miura-ori sheets connected along their crease lines, and it can maintain rigid foldability by following a set of simple geometrical compatibility rules  \cite{Schenk2013a}.  It has been shown that the rigid-foldable stacked Miura-ori can exhibit unique mechanical properties like negative Poisson's ratio \cite{Fang2016b}, shape morphing \cite{Li2015b,Sane2018a}, stiffness tuning by self-locking \cite{Fang2018}, and multi-stability (based on the spring hinge and rigid facet principle) \cite{Li2015c, Sengupta2018a}.   

This study uses a nonlinear bar-hinge numerical model (initially developed by Dr. Paulino's group) and experimental testing to investigate the stacked Miura-ori's stability characteristics as its facet varies from near rigid to soft.  We introduce two non-dimensional facet stiffness parameters and show that, when these parameters are in a certain range, the stacked Miura-ori unit cell can reach four stable states, two of which are unachievable if this cell's facet are rigid.   Moreover, the energy barrier between two of these four states is asymmetric, which means the stacked Miura-ori cell follows fundamentally different deformation paths and potential energy barriers as it switches between these two stable states.  Due to the asymmetric energy barrier, the origami cell requires a large force to be stretched from one stable state to the other but only requires a small compression force for the opposite switch.   Careful examination of the strain energy contribution reveals that this asymmetric energy barrier originates from the difference in facet stretching energy between extension and compression switch, and this is another evidence of the benefit from relaxing the rigid-folding conditions.  Moreover, one can assemble the unit cells into an architected structure while retaining the asymmetric multi-stability.  This study's results can provide new insights for exploiting the mechanics of non-rigid origami and foster new structural and material functionalities that traditional rigid origami cannot achieve.

In what follows, section 2 details the stacked Miura-ori unit cell's design and kinematics under rigid folding condition.  Section 3 explains the asymmetric multi-stability of the non-rigid unit cell, using numerical simulation and experimental testing.   Section 4 lays down the strategy of assembling the unit cells into a large-scale structure.  Section 5 concludes this paper with a summary and discussion.

\begin{figure}[t]
    \hspace{-0.8in}
    \includegraphics[scale=1.0]{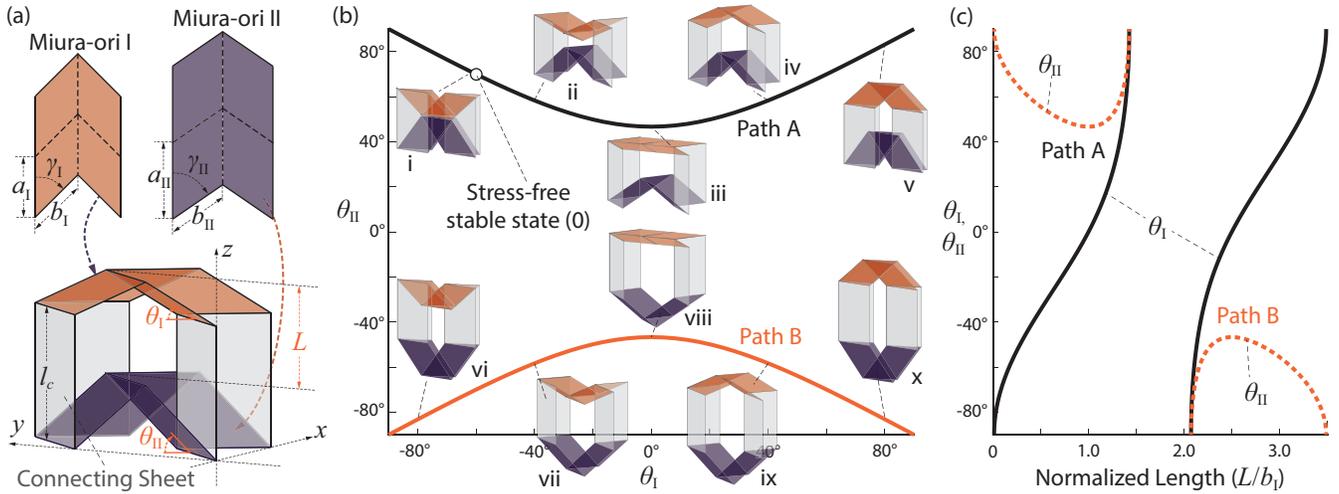}
    \caption{Design and rigid-folding kinematics of the stacked Miura-ori unit cell. (a) The design of the stacked Miura-ori unit cell consisting of two different Miura-ori sheets and accordion-shaped connecting sheets. The dihedral folding angles $ \theta_ \text{I} $ and $ \theta_ \text{II} $ are defined between Miura-ori sheets' facets and $ x-y $ reference plane. (b) The correlations between $ \theta_ \text{I} $ and $ \theta_ \text{II} $ according to Eq. \ref{eq:Rigid}, showing two disconnected kinematic paths. The subplots i-x illustrate the unit cell's shapes according to such rigid folding kinematics. (c) The same kinematic paths showed by folding angles with respect to normalized unit cell length.}
    \label{fig:Design}
\end{figure}

\section{Unit Cell's Design and Rigid Folding Kinematics}

Figure \ref{fig:Design}(a) illustrates the design of a stacked Miura-ori unit cell consisting of two different Miura-ori sheets and two accordion connecting sheets.  If the origami unit cell is rigid (aka. rigid facets and hinge-like creases), its deformation can be defined by two types of parameters \cite{Schenk2013a, Li2015b}. 
One is the geometric design parameters of the constituent origami sheets, including the connecting sheet's length $l_c$, Miura-ori's crease lengths $a_k$ and $b_k$, and sector angles $\gamma_k$ (here $k$ = I or II denotes Miura-ori sheet I and II, respectively).  To ensure geometric compatibility according to rigid-folding conditions, these design parameters should satisfy the constraints that $ b_\text{I} $= $ b_\text{II} $, $ a_\text{I} \cos \gamma_\text{I} = a_\text{II} \cos \gamma_\text{II} $. Here, we assume Miura-ori sheet II is bigger than sheet I in that $a_\text{II} > a_\text{I}$.  
The other type of parameter is the kinematic folding angles $\theta_k$, defined as the dihedral angles between Miura-ori sheet I and II's facets and the $x-y$ reference plane, respectively.  $\theta_k $ is denoted as positive if the corresponding Miura-ori sheet is ``above'' the $x-y$ plane and negative otherwise.  If this unit cell follows the rigid folding kinematics, there is no stretching, bending, or twisting deformation in the facets of Miura-ori and connecting sheets.  In this case, the two folding angels are not independent but rather follows the kinematic relationship in that  \cite{Fang2017}: 

\begin{equation} \label{eq:Rigid}
    \cos \theta_\text{I} \tan \gamma_\text{I} = \cos \theta_\text{II} \tan \gamma_\text{II}.
\end{equation}

Therefore, its rigid folding motion has only one degree of freedom.  The total length $L$ of the unit cell along the $z$-axis (the distance between the center vertices of the two Miura-ori sheets) is

\begin{equation}
L = a_\text{I} \sin \theta_\text{I} \sin \gamma_\text{I} - a_\text{II} \sin \theta_\text{II} \sin \gamma_\text{II} + l_c.
\end{equation}

Figure \ref{fig:Design}(b, c) elucidates the kinematics of a stacked Miura-ori unit cell under the rigid folding condition, and the corresponding design parameters are in Table \ref{tab:para}.   Since the Miura-ori sheets' folding angles $\theta_\text{I}$ and $\theta_\text{II}$ depend on each other (Eq. \ref{eq:Rigid}), the unit cell exhibits two \emph{disconnected} kinematic ``paths'' for rigid folding, which we label as Path A and B hereafter.  On Path A, the larger Miura-ori sheet II is always in a ``nested-in'' configuration ($\theta_\text{II}>0$), while on Path B, the sheet II is always ``bulged-out'' ($\theta_\text{II}<0$).  According to Eq. \ref{eq:Rigid}, a rigid-foldable unit cell cannot deform away from these kinematic paths and move from one path to the other.  This is because folding the larger Miura-ori sheets II between the nested-in and bulged-out configuration would incur facet deformations, which is not allowed under the rigid-folding conditions. 

On the other hand, if we relax the rigid-folding constraints and allow compliance in both Miura-ori sheets and connecting sheets, the origami facets behave like shell elements with complex deformation. Therefore, the non-rigid stacked origami cell can show richer deformation characteristics by deforming away from the two kinematic paths.  The following section demonstrates such non-rigid origami unit cells can obtain multi-stability properties that are unachievable for rigid origami cells.

\begin{table}[t]
\caption{Baseline design and constitutive material parameters of the stacked origami unit cell.  The material parameters are selected based on previous experiments \cite{Kaufmann2021a}. Notice that the crease folding stiffness in the connecting sheet $k_{f\text{C}}$ is significantly higher than other creases to ensure multi-stability \cite{Li2015b}.}
\label{tab:para}
\vspace{-0.05in}
\centering
    \scalebox{0.9}{
        \begin{tabular}{l l l l}
        \hline
         Geometry & Value & Material & Value \\
         \hline
         $ a_\text{I} $ & 20 mm                         & $ k_{f\text{M}} $ & 0.05 N/rad \\ 
         $ a_\text{II} $ & 25 mm                        & $ k_{f\text{CM}} $ & 0.05 N/rad \\
         $ b_\text{I} (= b_\text{II}) $ & 20 mm         & $ k_{f\text{C}} $ & $2 \sim 3$ N/rad \\
         $ l_c $ & 35 mm & $ k_s $                      & $ 10^4 $ N  \\
         $ \gamma_\text{I} $ & $ 45^\circ $             & $ \alpha $ & $ 10 \sim 10^5 $  \\
         $ \theta_\text{I}^0 $ & $ -60^\circ $          & $ \beta $ & $ 10 \sim 10^5 $  \\ 
         \hline
        \end{tabular}
    }
\end{table}

\section{Asymmetric Multi-Stability in a Non-Rigid Unit Cell}

\begin{figure}[t!]
    \hspace{-0.8in}
    \vspace{-0.05in}
    \includegraphics[scale=.95]{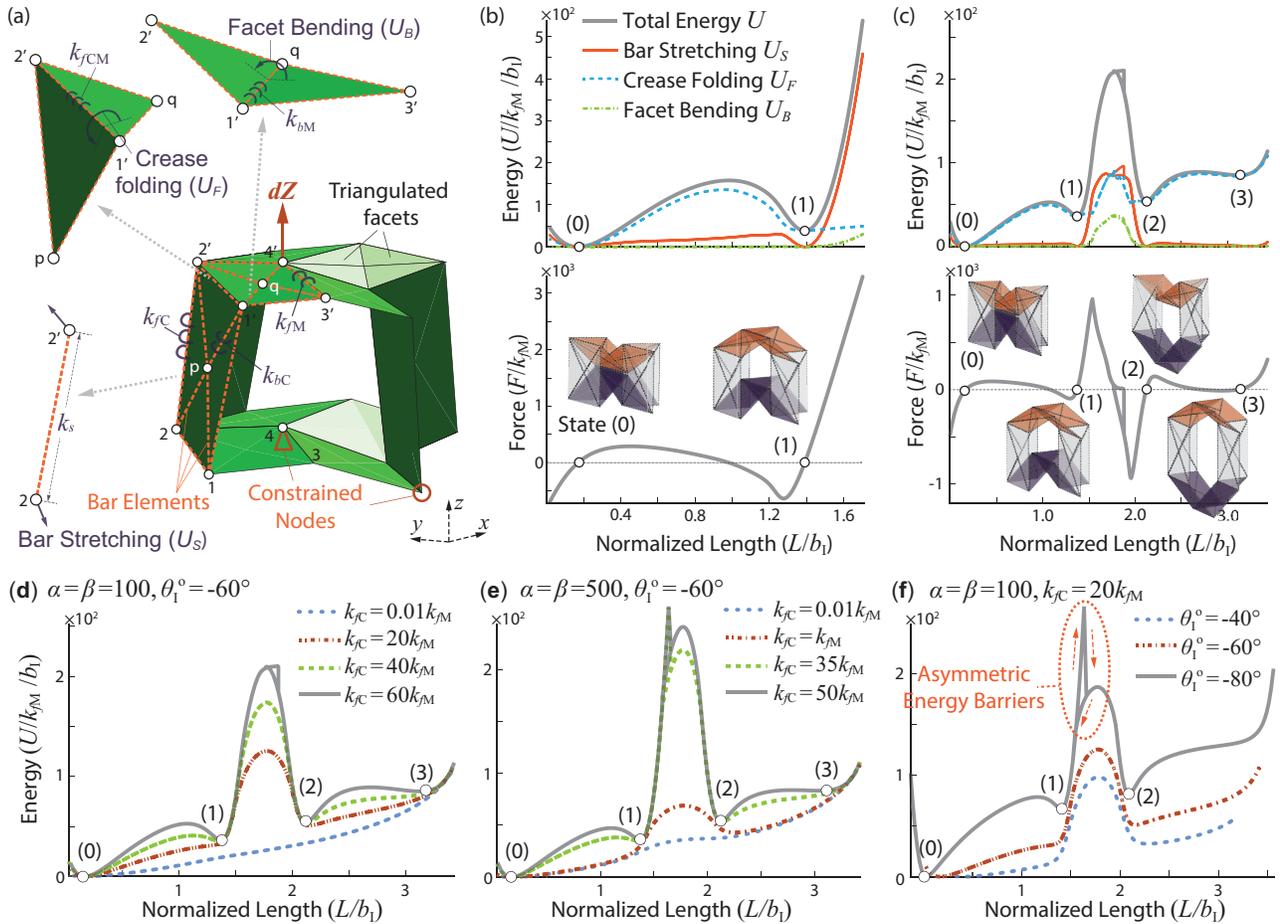}
    \caption{Bar-hinge model and the multi-stability behavior of the non-rigid unit cell. (a): The N5B8 scheme in the unit cell based on the bar-hinge approach. The detailed model formulation could be found in Appendix A. The load and boundary conditions of the numerical simulations are highlighted: the red triangle denotes a fully constrained node with no displacements, and red circle denotes a partially constrained node that can displace only in $y$ and $z-$direction.  We apply controlled $z$-directional displacement $dZ$ at node $4'$. (b): The potential energy landscape (top) and corresponding force displacement curve (bottom) of a near rigid unit cell. Here, $ \alpha=10^5, \beta=10^5 $, $ \theta_\text{I}^0 = -60\circ $ and $ k_{f\text{C}}/ k_{f\text{M}} = 60 $.  (c): The energy landscape and force-displacement curve of a non-rigid unit cell, showing the two additional stable states (2) and (3).  Here, $ \alpha = 100 $, $ \beta = 100 $, $ \theta_\text{I}^0 =60\circ $ and $ k_{f\text{C}}/ k_{f\text{M}} = 60 $. The subplots in (b) and (c) are bar-hinge model's prediction of the folding configurations at different stable states.  (d-f): The correlation between non-rigid unit cell's multi-stability and it design parameters, including (d) crease stiffness ratio $ k_{f\text{C}}/ k_{f\text{M}}$, (e) facet stiffness $\alpha$, $\beta$, and (f) stress-free folding angle $ \theta_\text{I}^0 $.}
    \label{fig:Stability}
\end{figure}

To accommodate and analyze facet deformations in a stacked origami unit cell, we adopt the nonlinear bar-hinge modeling approach that discretizes the continuous origami structure into a pin-jointed truss-frame mechanism \cite{Schenk2013a, Li2015b, Gillman2018e, Liu2018, Liu2019a}.   This model uses stretchable bar elements to represent the origami crease and diagonalize the facets, and it also adds rotational stiffness between the triangles defined by these trusses to estimate crease folding and facet bending stiffness (Figure \ref{fig:Stability}(a)).  This reduced-order model can analyze different origamis' primary deformations without incurring expensive computational costs like finite element simulations.  
This study uses the open-source MERLIN2 software to simulate the unit cell's multi-stability under displacement control \cite{Liu2018}.  In particular, we apply the ``5-Nodes-8-Bars'' (N5B8) triangulation scheme that adds a Steiner point at the intersection of the two diagonals of a quadrilateral facet (e.g., point $p$ in the $1-2-2'-1'$ facet in (Figure \ref{fig:Stability}(a)). The N5B8 scheme allows the discrete truss-frame system to capture the more realistic doubly-curved out-of-plane deformations and isotropic in-plane deformations in the thin facets, potentially yielding a higher accuracy \cite{Filipov2017}.

We assume the bar-hinge system's potential energy is conservative and only a function of the current deformation configuration. The total potential energy $U$ of the unit cell is a summation of bar stretching energy ($ U_S $), crease folding energy ($ U_F $), as well as facet bending/twisting energy ($ U_B $) so that 

\begin{equation}\label{eq:Energy}
   U=U_S + U_F + U_B. 
\end{equation}

Here, the bar stretching energy represents the in-plane stretching and shearing of the origami facets. The principle of stationary potential energy, which accounts for both constitutive material properties and nonlinear finite deformation, can be used to solve the unit cell's mechanical response.   A summary of the fundamental formulation of this nonlinear bar-hinge model is available in Appendix A, and interested readers can refer to the relevant literature for further details \cite{Liu2018, Liu2019a, Filipov2017}.

In the bar-hinge framework, the non-rigid origami's deformation is approximated based on the bar stretching rigidity $k_s$ that contributes to $U_S$ in Eq. (\ref{eq:Energy}), torsional crease folding stiffness $k_f$ that contributes to $U_F$, and facet bending stiffness $k_b$ that contributes to $U_B$.  For clarity, we further categorize the crease folding stiffness into three different groups: $ k_{f\text{M}} $ is the torsional stiffness \emph{per unit length} of the creases in Miura-ori sheets I and II (e.g., along creases $3-4$ or $3'-4'$ in Figure \ref{fig:Stability}(a)).  $ k_{f\text{CM}} $ is the torsional stiffness of the creases between the connecting sheets and the Miura-ori sheets (e.g., along creases $1-2$ or $1'-2'$).  $ k_{f\text{C}} $ is the torsional stiffness along the creases in the connecting sheets (e.g., along crease $2-2'$).  The facet bending stiffness $k_b$ can also be divided into two groups: $k_{b\text{M}}$ for the Miura-ori's facets (e.g., along $1'-q$) and $k_{b\text{C}}$ for the connecting sheet's facets (e.g., along $1'-p$).   Moreover, we assume all these stiffness coefficients are constant so that the non-linearity originates from the finite-amplitude deformation during folding.   

In this study, the ratio between the facet bending stiffness and crease folding stiffness ($ k_b / k_f $) is crucial because it determines whether the origami follows the rigid folding conditions \cite{Liu2017e}.  A large $ k_b / k_f $ ratio (e.g., $ 10^5 $) indicates that facets are near rigid with minimal deformation so that the rigid-folding kinematics discussed in the previous section would dominate.  In contrast, a small $ k_b / k_f $ ratio (e.g., $ 10 $) indicates a non-rigid origami where the facet deformations are significant.  Therefore, we introduce two dimensionless parameters to describe the relative facet rigidity of the Miura-ori sheets and connecting sheets in that

\begin{align}
    \alpha & = \frac{k_{b\text{M}}}{k_{f\text{M}}} \\
    \beta  & = \frac{k_{b\text{C}}}{k_{f\text{C}}}
\end{align}

Based on $\alpha$ and $\beta$, we can quantitatively analyze the multi-stability of the unit cell either with or without the rigid folding assumption. Unless noted otherwise, the unit cells in the subsequent studies have an initial folding angle $ \theta_\text{I}^0=-60^\circ, \theta_\text{II}^0=70^\circ $, where all creases and facets are stress-free without any deformations.   This initial configuration is on kinematic Path A as highlighted in Figure \ref{fig:Design}, and we denote it as the stable state (0) hereafter. The baseline values of the bar-hinge stiffness coefficients are also summarized in Table \ref{tab:para}.

\subsection{Bi-Stability in a Rigid Unit Cell}
We first consider the case of rigid-foldable origami by setting $ \alpha $ and $\beta $ to a high value $ 10^5 $ in the bar-hinge model.   As a result, as we stretch the unit cell along the $z$-axis under displacement control, it will switch from the initial stable state (0) to a new stable state (1) \emph{near} the same kinematic path A (Figure \ref{fig:Stability}(b)).  During this switch, the crease folding energy $U_F$ is dominant compared to bar stretching $U_S$ and facet bending energy $U_B$. However, as we continue to stretch the unit cell beyond state (1), its reaction force will increase significantly as the bar-stretching energy becomes dominant (here, the reaction force is the first variation of strain energy with respect to the unit cell length, i.e., $ F = \partial U /\partial L $).   As a result, no other stable states are achievable.  Such bi-stability is fundamentally the same as the multi-stability studied in other rigid origamis in that it originates from the combination of torsional stiffness from the hinge-like creases and nonlinear kinematics of rigid folding \cite{Fang2017, Baharisangari2019c}.  

\subsection{Multi-Stability in a Non-Rigid Unit Cell}
If we relax the rigid-folding constraints and allow compliance in the origami facets, the unit cell can reach more stable stables.  Figure \ref{fig:Stability}(c) illustrates the potential energy landscape of the non-rigid unit cell when $ \alpha=\beta=100 $ and $ k_{f\text{C}}/ k_{f\text{M}}= 60 $.  One can see that the compliance in origami facets allows the unit cell to deform away from kinematic Path A and reach Path B, exhibiting two new stable states (2) and (3).  As the unit cell switches between state (0) and (1), or between state (2) and (3), the crease folding energy $U_F$ constitutes the majority of total strain energy.  So the unit cell roughly follows the rigid-folding kinematics between these two pairs of stable states because each pair is roughly on the corresponding kinematic path. However, when the unit cell switch between state (1) and (2), the bar stretching energy and facet bending energy play the dominant role, indicating that the unit cell has to violate the rigid-folding kinematics to deform between these two states. Since the facet stretching and bending give more resistance than crease folding, the magnitude of the energy barrier is the highest between stable states (1) and (2).  A similar trend is evident in the corresponding reaction force plot as shown in Figure \ref{fig:Stability}(c). The magnitude of the reaction forces between stable states (1) and (2) are significantly higher than other switches. 

The occurrence of these four different stable states are directly related to the geometry and material properties of stacked origami.  One can use the nonlinear bar-hinge models to examine such relationships: 

\begin{itemize}
    \item {\bf Crease folding stiffness ratio $ k_{f\text{C}}/ k_{f\text{M}}$:} Generally speaking, multiple potential energy wells, which are the defining characteristics of a multi-stable system, start to show up as the connecting sheets' crease folding stiffness ($ k_{f\text{C}} $) is sufficiently larger than that of the Miura-ori sheets ($ k_{f\text{M}} $).  Interestingly, if the connecting sheets' crease folding stiffness is only moderately higher than the Miura-ori sheets (e.g., $k_{f\text{C}} / k_{f\text{M}}=20 $ in Figure \ref{fig:Stability}(d), the unit cell is only bi-stable at state (0) and state (2).  This means that there is only one stable state on each kinematic path, and the energy barrier from the non-rigid folding separates these two states.  As the $k_{f\text{C}} / k_{f\text{M}} $ ratio increases, the stable state (1) emerges near the kinematic path A so that the switching sequence (0) $ \leftrightarrow (1) \leftrightarrow (2) $ is possible under cyclic loads.  Finally, the stable state (3) appears as the $k_{f\text{C}} / k_{f\text{M}} $ increases further. 
    
    \item {\bf Facet stiffness $\alpha$, $\beta$:} Figure \ref{fig:Stability}(e) illustrates the effect of the facet stiffness on unit cell's multi-stability properties. As the relative rigidity of facets $ \alpha, \beta $ increases, the required $k_{f\text{C}} / k_{f\text{M}} $ ratio for achieving the similar multi-stability is reduced.  However, if the facets are rigid enough, switching between state (1) and (2) is no longer possible.  The criterion for obtaining the multi-stability on both kinematic paths is $ \alpha, \beta < 750$ if other design and material parameters follows Table \ref{tab:para}. 
    
    \item {\bf Stress-free folding angle $ \theta_\text{I}^0 $:}  The effect of  $\theta_\text{I}^0$ is shown in Figure \ref{fig:Stability}(f). As $ \theta_\text{I}^0 $ deviates further away from $ 0^\circ $, the required $k_{f\text{C}} / k_{f\text{M}} $ for achieving the same multi-stability is reduced, meaning that it is easier to achieve multi-stability.
    
\end{itemize}

\subsection{Asymmetric Energy Barrier}

The parametric studies of the non-rigid unit cell's multi-stability also revealed an interesting phenomenon:  Under certain combinations of design and material parameters, the unit cell follows different potential energy landscapes as it is stretched or compressed between the stable states (1) and (2) (highlighted in Figure \ref{fig:Stability}(f)).  To examine the origin of such asymmetric energy barrier in detail, we conducted additional simulations with $ \alpha=500 $ and $ k_{f\text{C}} / k_{f\text{M}}=40$ (other parameters follow Table \ref{tab:para}).  According to the previous subsection's results, the origami unit cell with such designs should possess three stable states (0), (1), and (2).

\begin{figure}[t]
    \hspace{-0.7in}
    \includegraphics[scale=1.0]{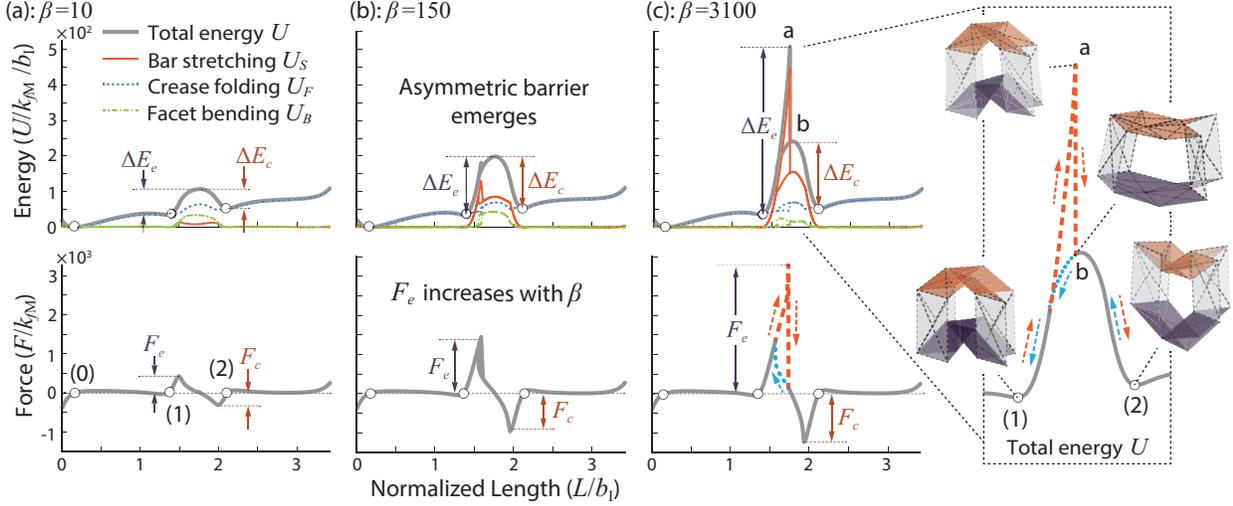}
    \caption{Origin of the asymmetric energy barrier.  In this case, $ \alpha=500 $ and $k_{f\text{C}} / k_{f\text{M}}=40 $. The energy landscapes (first row), and the corresponding reaction forces (second row) change fundamentally as the connecting sheets' facet stiffness $\beta$ increases: (a) $ \beta = 10 $, (b) $ \beta = 150 $, and (c) $ \beta = 3100 $. If $ \beta > 3100 $, the origami unit cell is effectively rigid and cannot deform from state (1) to (2). Energy contributions from three deformation modes are highlighted.  The subplot on the right details the unit cell's deformation between states (1) and (2).}
    \label{fig:Asymmetry_Origin}
\end{figure}

We first consider a case that the facet stiffness of the connecting sheet is relatively small, i.e., $ \beta =10 $.  This assumption corresponds to a ``soft'' connecting sheet in the unit cell.   Figure \ref{fig:Asymmetry_Origin}(a) shows the corresponding energy landscape and force-displacement curves calculated by the nonlinear bar-hinge model.  One can see that the unit cell follows roughly the same energy and force-displacement curves between stable states (1) and (2), so asymmetry in the energy barrier does not exist.

As the facet stiffness of the connecting sheet increases, the unit cell's behaviors change significantly. In particular, if $ \beta $ is large enough (e.g., $ \beta = 150 $ in Figure \ref{fig:Asymmetry_Origin}(b)), the potential energy landscape and force-displacement curves start to ``split'' between states (1) and (2), creating an asymmetric energy barrier.  Such asymmetry becomes more prominent as $\beta$ continues to increase (e.g.,  $ \beta = 3100 $ in Figure \ref{fig:Asymmetry_Origin}(c)).  That is, if the unit cell is monotonically extended from the stable state (1) under displacement control, it will deform by following an energy path until reaching a point '$ a $' shown in Figure \ref{fig:Asymmetry_Origin}(c), then it will rapidly ``jump'' to the other energy path at position '$ b $' before reaching stable state (2).  During this jump, the internal folding configuration of the unit cell and the reaction force will change significantly even though the total unit cell length changes little.  On the other hand, if the unit cell is compressed monotonically from the stable state (2), it will follow another energy path to the state (1) without any obvious jump.  

A careful examination of the unit cell's 3D deformation and its potential energy composition can reveal the physics underpinning such differences between extension and compression.  At state (1), the larger Miura-ori sheet II is in the ``nested-in'' configuration.  As one extends the unit cell from this state, the rigid folding kinematic constraints are violated, and Miura-ori sheet II would significantly stretch the connecting sheet as it tries to deform to the ``bulged-out'' configuration at the stable state (2).  As a result, the bar stretching energy $U_S$ dominates right before the unit cell reaches point '$a$.'   However, such connecting facets' stretching doesn't occur when the unit cell is compressed from the stable state (2) back to (1).

As a result, when the facet stiffness of the connecting sheet increases, the maximum energy barrier for deforming from the stable state (1) to (2) increases significantly ($ \Delta E_e $ in the first row of Figure \ref{fig:Asymmetry_Origin}), while the energy barrier of the opposite switch does not increase as much ($ \Delta E_c $).  Similarly, the external force required to extend the unit cell from the stable state (1) to (2) ($ F_e $ in the second row of Figure \ref{fig:Asymmetry_Origin}) also increases significantly.  However, the required compression force for the opposite switch $ F_c $ does not increase as much (i.e., $ F_e \gg F_c $).  However, it is worth noting that if the connecting facet stiffness is too high, the unit cell becomes effectively rigid, so it can no longer switch between states (1) and (2). In the combination of the designed parameters in Table \ref{tab:para}, the criterion for the switches is $ \beta < 3100 $ when $ \alpha = 500 $.

Therefore, to obtain asymmetrical energy barrier, the connecting sheet's facets need to be reasonably stiff (e.g., $ 150 < \beta  < 3100 $ in the case shown in Figure \ref{fig:Asymmetry_Origin}). The asymmetry of multi-stability are the strongest at the ``threshold'' between non-rigid and rigid origami conditions.  

Moreover, the asymmetry of multi-stability is also related to the origami unit cells' geometry and material design parameters.  To this end, we conduct further case studies:

\begin{figure}[t]
    \includegraphics[scale=1.0]{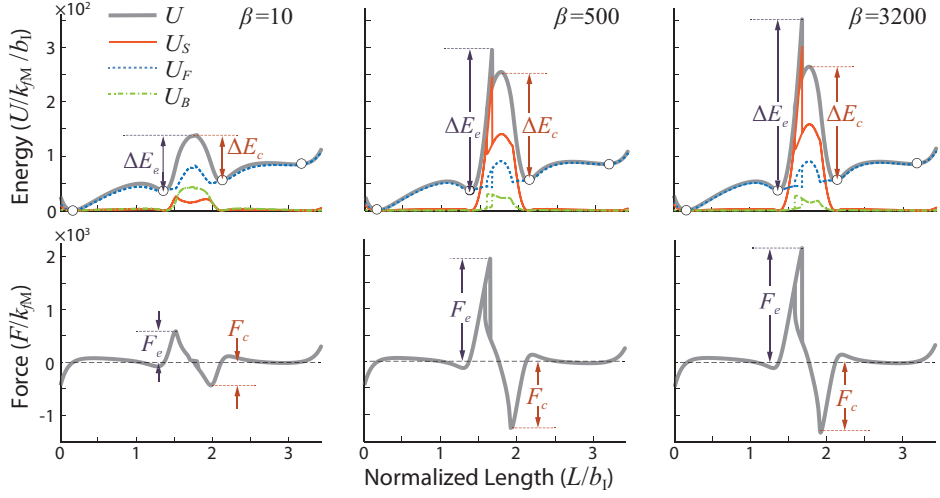}
      \centering
    \caption{Effect of $k_{f\text{C}} / k_{f\text{M}}$ on the asymmetric energy barrier. In this case, $ \alpha=500 $ and $k_{f\text{C}} / k_{f\text{M}}=60 $. When $ \beta =500 $, the potential energy landscape and force-displacement curves start to ``split'' between states (1) and (2), creating an asymmetric energy barrier. When $ \beta > 3200 $, the unit cell becomes too rigid to deform from state (1) to (2).}
    \label{fig:AsymmetryPara1}
\end{figure}
 
\begin{itemize}
    \item {\bf Crease folding stiffness ratio $ k_{f\text{C}}/ k_{f\text{M}}$:} Figure \ref{fig:AsymmetryPara1} summarizes the energy landscape and force-displacement curves when the crease stiffness ratio $k_{f\text{C}} / k_{f\text{M}}$ is higher at 60 (this ratio is 40 in the previous case study in Figure \ref{fig:Asymmetry_Origin}).  Surprisingly, although there are four stable states exist in this case rather than three, the asymmetry of multi-stability becomes weaker with a lower ratio of $ \Delta E_e / \Delta E_c $.  
    
    \item {\bf Stress-free folding angle $ \theta_\text{I}^0 $:}  Compared to the crease stiffness ratio, the initial stress-free configuration of the unit cell seems to exert a more significant impact on the asymmetry of multi-stability (Figure \ref{fig:AsymmetryPara2}).  Comparing to $ \theta_\text{I}^0 = -60\circ $ in Figure \ref{fig:Asymmetry_Origin}, one can conclude that as the  $ \theta_\text{I}^0 $ deviates further away from $ 0^\circ $ in which the unit cell is more ``compact'' initially, the asymmetry in the energy barrier becomes more prominent.
    
\begin{figure}[ht]
    \includegraphics[scale=1.0]{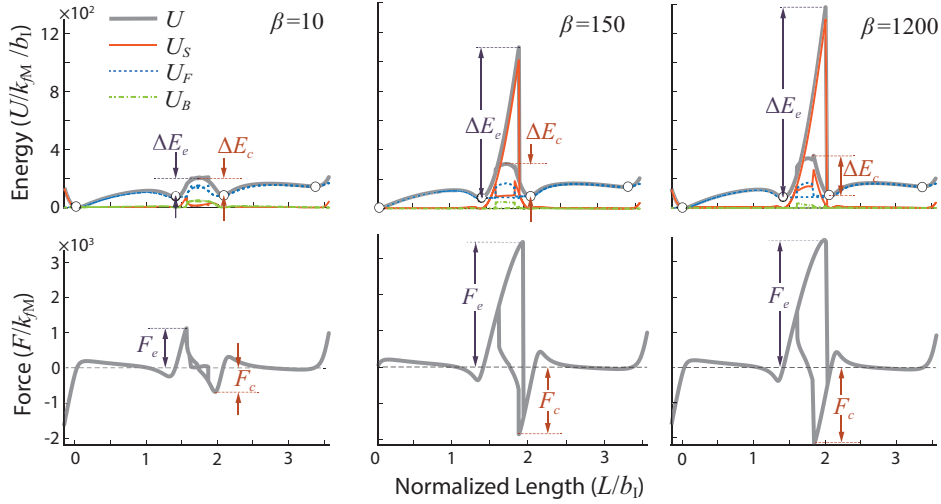}
      \centering
    \caption{Effect of $ \theta_I^\circ $. In this case, $ \alpha=500 $, $ \theta_I^\circ=-80^\circ $. When $ \beta =150 $, the asymmetry in the energy barrier has been pretty prominent. When $ \beta > =1200 $, the unit cell becomes too rigid to deform from state (1) to (2).}
    \label{fig:AsymmetryPara2}
\end{figure}

\begin{figure}[ht]
    \includegraphics[scale=1.0]{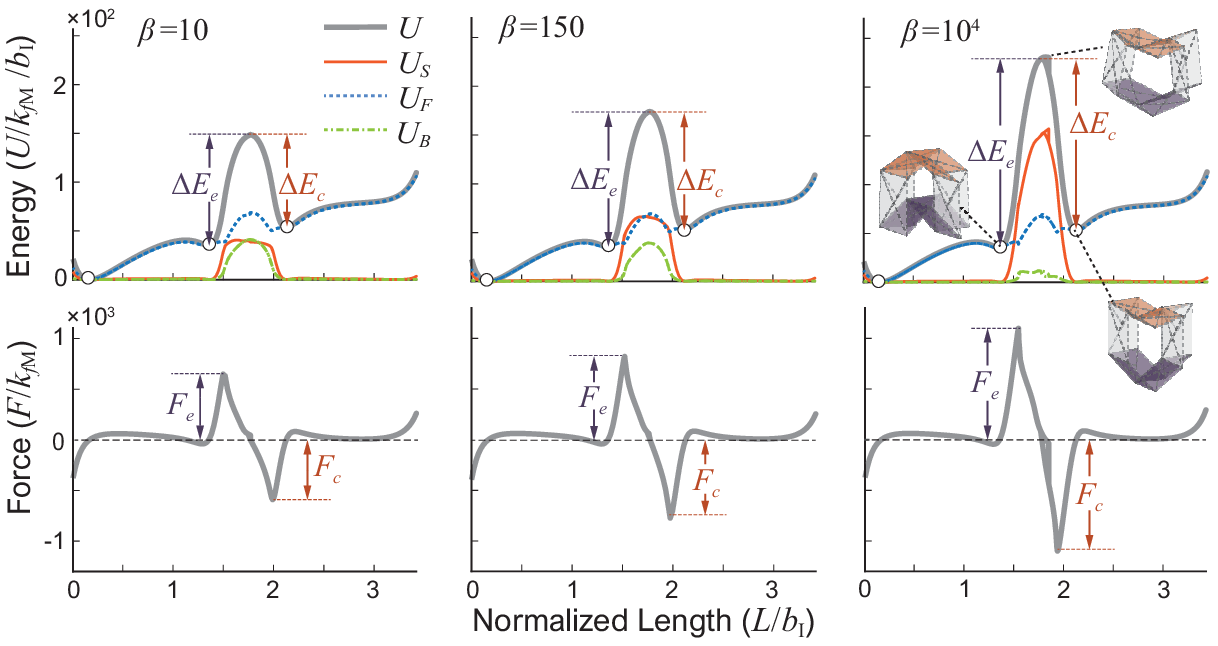}
      \centering
    \caption{Effect of unit cell's rigidity on the asymmetry of multi-stability in unit cell. In this case, $ \alpha=100$, $k_{f\text{C}} / k_{f\text{M}}=40 $ and $ \theta_I^\circ=-60^\circ $. Interestingly, there is no asymmetric energy barrier because the highly \emph{soft} Miura-ori's facets bend and stretch significantly in both extension and compression between states (1) and (2) so that no significant facets' stretching exists in the connecting sheet even though $\beta$ is high at $10^4$. The subplots show the unit cell's deformation between states (1) and (2).}
    \label{fig:AsymmetryPara3}
\end{figure}

    \item {\bf Origami Facet stiffness $\alpha$, $\beta$:}  Interestingly, we find that there is no asymmetry in multi-stability if the Miura-ori's facets are too ``soft'' (aka. $\alpha=100$ in Figure \ref{fig:AsymmetryPara3}).  The subplots in this figure show that the Miura-ori's facets bend and stretch significantly in both extension and compression between states (1) and (2).  As a result, it could not induce significant facet stretch in the connecting sheet and create asymmetry, even if the facet rigidity of the connecting sheet is very high ($\beta = 10^4$). 
    
\end{itemize}

Based on these additional parametric studies, one can conclude that there should exist an optimal set of design parameters to achieve significant asymmetry in multi-stability. Careful designs are necessary according to potential application requirements.

\begin{figure}[b!]    
    \hspace{-0.8in}
    \includegraphics[scale=1.0]{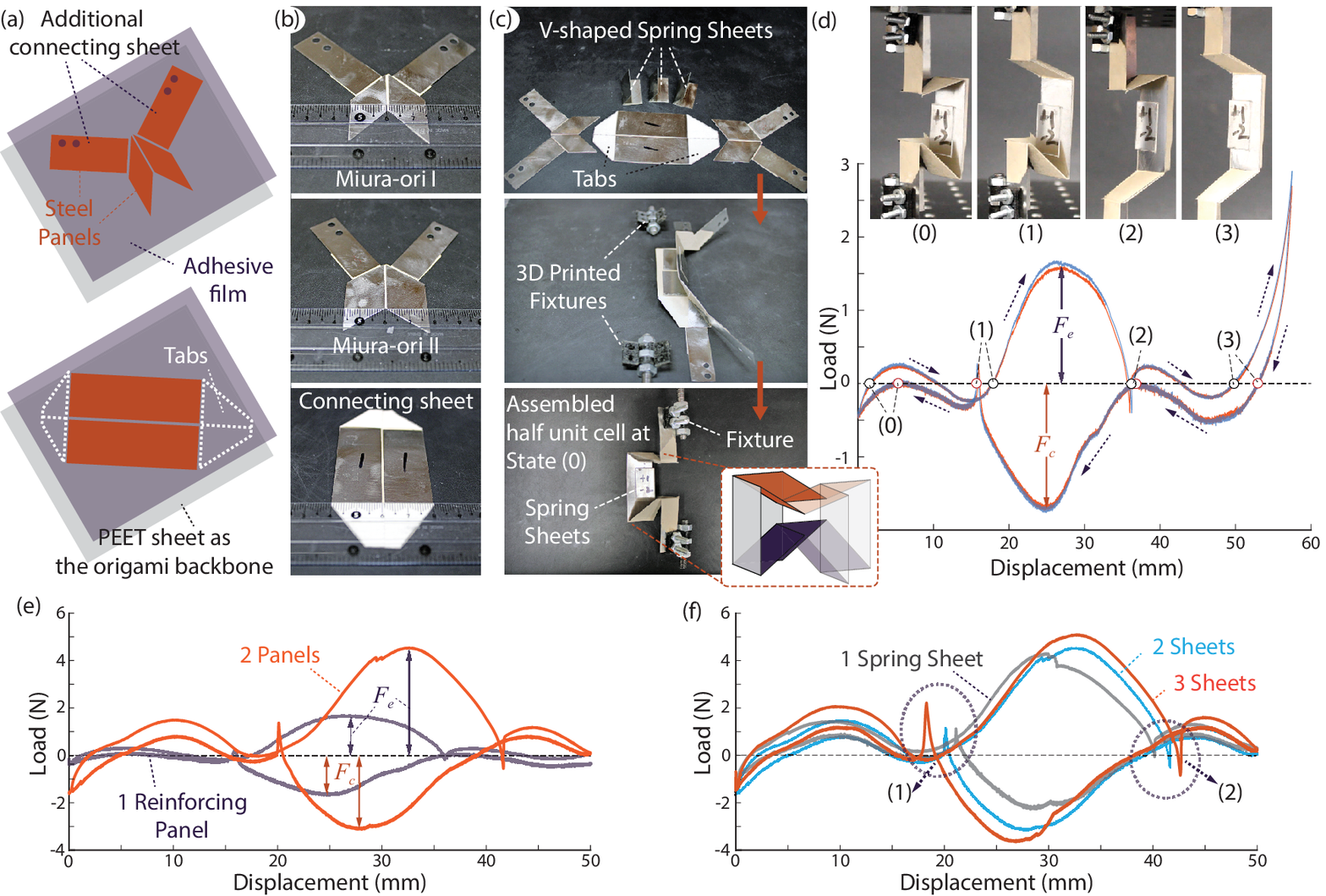}
    \caption{Fabrication and testing of a half-unit cell prototype. (a) Layered construction method showing the PEET backbone, adhesive film, and steel panels for reinforcing the facets. Additional connecting sheets accommodate 3D printed fixtures that can attach the cell to the universal tester machine. The tabs are used to connect Miura-ori sheets and connecting sheets. (b) Finished Miura-ori sheets and connecting sheet. (c) The assembly of half unit cell. The insert drawing explains the corresponding unit cell prototype at stable state (0). (d) Experimentally measured force-displacement curves, showing the switching between four stable states and asymmetric energy barrier between states (1) and (2).  Results from two loading cycles are plotted with excellent repeatability. (e, f) Additional experiment showing the correlation between facet stiffness, crease stiffness ratio, multi-stability, and asymmetry properties. These results are the average of two loading cycles.} 
    \label{fig:fabrication}
\end{figure}

\subsection{Experiment Assessments}

To experimentally assess the asymmetric multi-stability in the proposed designs, we fabricated and tested unit cell prototypes as shown in Figure \ref{fig:fabrication}(a-c).  Only half of the unit cell is fabricated to avoid unnecessary fabrication complexities.  First, we cut and fold thin plastic sheets (PEET, 0.1 mm thickness) into an origami ``backbone'' and paste waterjet-cut shim stock to stiffen the facets (316 stainless steel, 0.25 mm thickness).  The plastic origami backbone and steel panels are bonded by double-sided adhesive films.  The gap between the reinforcing steel facet (or the width of plastic crease lines) is roughly 1mm.  To increase the $k_{f\text{C}}/ k_{f\text{M}}$ ratio and achieve multi-stability, we attach pre-bent, V-shaped spring steel sheets along the connecting sheet's crease lines.   Finally, the assembled prototypes are annealed at 60$^\circ$C for 90 minutes to relieve the residual stress in the plastic origami backbone due to folding.

Figure \ref{fig:fabrication}(d) shows the measured force-displacement curves of the half-unit cell prototype under cyclic loading (ADMET eXpert 5061 with 25 lbs load cell). In each cyclic loading cycle, the unit cell prototype is stretched from the stress-free stable state (0) to beyond state (3) and then compressed back to below state (0).  The unit cell prototype exhibits noticeable plastic deformation in the first loading cycle but then shows excellent repeatability in subsequent cycles.  The experiment results in this figure confirm the existence of four stable states and the predicted switching sequence among them under displacement control.  Moreover, they directly indicate the occurrence of asymmetric energy barriers.  That is, the unit cell follows similar force-displacement curves while switching between states (0) and (1) or between states (2) and (3) (albeit some hysteresis behaviors).  However, it follows fundamentally different force-displacement curves between the stretch from states (1) to (2) and compression from (2) to (1).  

To qualitatively test the correlation between the asymmetry of multi-stability and unit cell design, we conducted further experiments. Figure \ref{fig:fabrication}(e) shows the results from two half-unit cell prototypes with 1 and 2 reinforcing steel panels on their connecting sheet's facets.  Both cells have 3 V-shaped spring steels on the connecting sheet's creases, so they show the same $k_{f\text{C}}/ k_{f\text{M}}$ ratio but different $\beta$ values. While the unit cell with 1 reinforcing panel (low $\beta$) does not show a strong asymmetric energy barrier with $F_e/F_c \approx 1$, the other cell with two reinforcing panels (higher $\beta$) shows a stronger asymmetry with $F_e/F_c \approx 1.4$.  This trend is consistent with the numerical simulations.  We also tested another unit cell prototype with an even higher $beta$ with 3 reinforcing panels on the connecting sheet's facets.  However, this unit cell can no longer be switched from states (1) to (2), indicating that this unit cell is too ``rigid'' and loses the multi-stability all together.  Again, this trend is consistent with numerical simulations in the previous subsection.
    
Figure \ref{fig:fabrication}(f) summarizes the force-displacement curves of three prototypes with 1, 2, and 3 V-shaped spring steel sheets in their connecting sheets.  All these cells have 2 reinforcing steel panels in their connecting sheet's facets. Notice that additional V-shaped spring steel sheets increases both $ k_{f\text{C}}/ k_{f\text{M}}$ ratio and $\beta$. The unit cell prototype with only 1 V-shaped spring steel can not reliably reach one stable states (1) or (2). On the other hand, more spring steel sheets give more prominent existence of state (1) and (2). This trend is also consistent with earlier numerical simulations.

\section{Assembling into a Cellular Solid}

\begin{figure}[t!]    
    \centering
    \includegraphics[scale=0.96]{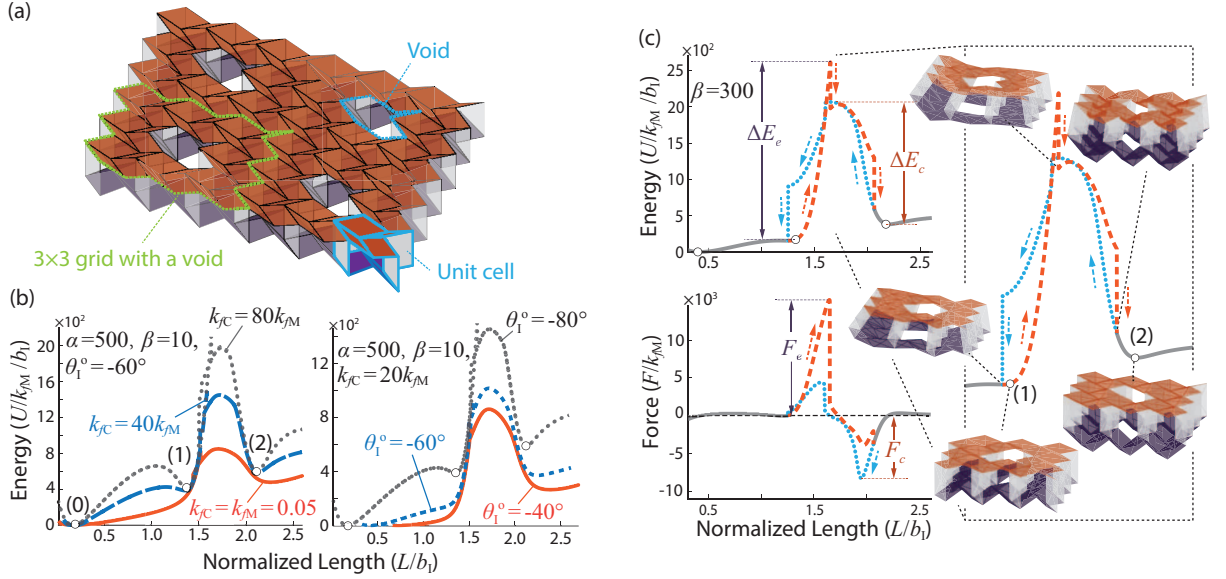}
    \caption{Construction of Cellular Solids. (a) The strategy to construct a cellular structure. The  3$\times$3 cellular grid with one center void is highlighted. (b) Potential energy landscape of the 3$\times$3 cellular grid based on different design parameters. The multi-stability exists when the solid has a high ratio of $k_{f\text{C}} / k_{f\text{M}}=40$ or the initial folding angle deviates away from $0\circ$. This is similar to the previous results from a single unit cell. (c) Response of a 3$\times$3 grid with $\beta=300$, $\theta_\text{I}^o=-60^\circ$, and $k_{f\text{C}} / k_{f\text{M}}=40$. The asymmetrical energy barrier exists when the connecting sheet's facets are reasonably stiff (e.g., $ 10 < \beta  < 300 $. The plot on the right details the element's deformation between states (1) and (2) from the non-linear bar-hinge model.}
    \label{fig:Cellular}
\end{figure}

While the previous sections focus on a single unit cell, this section discusses the multi-stability of an origami cellular consisting of multiple unit cells.   A unique challenge here is that when unit cells are assembled along the $x$ and $y-$direction, they will reinforce the rigid-folding kinematics to each other because of their periodic nature.  As a result, a ``closely packed'' unit cell assembly could not exhibit multi-stability and asymmetric energy barrier even if their facets are compliant. Therefore, we introduce a cellular assembly scheme with ``voids'' as shown in Figure \ref{fig:Cellular}(a).   These voids introduce additional kinematic freedoms to accommodate the non-rigid origami deformation of the unit cells, thus retaining the asymmetric multi-stability.  A bar-hinge model is constructed on the fundamental element of these cellular structure --- a 3$\times$3 cellular grid with a void in the center--- and the corresponding simulation results are provided.  As shown in Figure \ref{fig:Cellular}(b), one can find that the multi-stability exists when $ k_{f\text{C}} / k_{f\text{M}} = 40 $ with $ \theta_I^o=-60^\circ $, or when $ k_{f\text{C}} / k_{f\text{M}} = 20 $ with $ \theta_I^o=-80^\circ $.  In the case of $\alpha=500$, $k_{f\text{C}} / k_{f\text{M}}=40 $ and $ \theta_I^\circ=-60^\circ $, the energy barriers for switching between two stable states in the proposed 3$\times$3 cellular grid, especially between states (1) and (2), are nearly 10 times larger than that of a single unit cell discussed in Figure \ref{fig:Asymmetry_Origin}(a).  Figure \ref{fig:Cellular}(c) elucidates that the required facet stiffness of the connecting sheet for achieving asymmetric energy barrier is also much smaller than that of a single unit cell in Figure \ref{fig:Asymmetry_Origin}(c). This is because the assembly of unit cells along the $x$ and $y-$direction induces a strengthened rigid-folding constraints.  Moreover, the asymmetry of multi-stability in the proposed 3$\times$3 cellular grid becomes weak with a lower ratio of $ \Delta E_e / \Delta E_c $.

Therefore, a reasonably stiff connecting sheet facet can also create the asymmetric energy barrier in an assembled cellular solid.  Like in the unit cell, careful design is necessary to select the optimal design for specific application requirements.

\section{Summary and Conclusion}

Via numerical simulations and experimental testing, this study examines the multi-stability and asymmetric energy barrier between stable states that emerge from intentionally relaxing the rigid folding conditions in a stacked origami cellular structure.  The unit cells in such a structure combine two different Miura-ori sheets and accordion-shaped connecting sheets.  By introducing two non-dimensional parameters $ \alpha $, $ \beta $ to describe the origami facets' relative rigidity, we conduct a quantitative analysis of the unit cell's multi-stability based on the nonlinear bar-hinge approach.  That is, high $ \alpha $ and $ \beta $ values indicate stiff origami facets that reinforce the rigid-folding kinematics.  In contrast, low $ \alpha $ and $\beta $ represent more compliant origami facets that can accommodate additional deformation patterns.  The simulation results show that only two stable states exist in the unit cell if it follows the rigid origami kinematics; however, two more stable states are reachable if the origami facets become sufficiently compliant.  Moreover, the switches between the stable states (1) and (2) --- which are on two different kinematic paths --- shows an asymmetric energy barrier, meaning that the unit cell follows fundamentally different deformation paths when it extends from the state (1) to (2) compared to the opposite compression from (2) to (1).   As a result, the reaction force required for extension between these two states is also significantly higher than compression.  A close examination of the unit cell's overall potential energy reveals that the facet stretching plays a significant role in such asymmetry, and the overall design of such origami unit cells can be exploited to fine-tune the multi-stability behaviors.   Finally, the asymmetric multi-stability are validated in experiments, and a strategy to assemble such unit cells into a cellular structure is put forward.  By showing the benefits of exploiting facet compliance, this study could foster multi-functional structures and material systems that traditional rigid origami cannot create.

\section{Acknowledgments}

The authors acknowledge the partial support from the National Science Foundation (Award \# CMMI-1751449 CAREER) and Clemson University (via startup funding and the CECAS Dean’s Faculty Fellow Award) and China Scholarship Council.

\appendix
\section{Bar-Hinge Model Formulation Fundamentals}

The bar-hinge approach discretizes the continuous origami structure into a pin-jointed truss-frame mechanism \cite{Gillman2018e, Li2015b, Schenk2013a}.  It uses stretchable bar elements to represent the origami crease and diagonalize the facets. To estimate crease folding and facet bending, this model adds rotational stiffness between the triangles defined by these trusses. 
As illustrated in Figure \ref{fig:Stability}(a)), we use the N5B8 triangulation scheme so that the discrete model can potentially yield higher resolution than the previous N4B5 scheme \cite{Filipov2017}.

In the bar-hinge system, the total potential energy is the summation of bar stretching energy ($ U_S $), crease folding energy  ($ U_F $), as well as facet bending/twisting energy ($ U_B $), as shown in Eq. \ref{eq:Energy}. 
%
%
The critical step in formulating the unit cell's mechanics model is to obtain the tangent stiffness matrix (aka. a second-order approximation of the potential energy $U$) as \cite{Filipov2017,Liu2017e}

\begin{equation}
\label{eq:K}
\mathbf{K}=\mathbf{K}_S+\mathbf{K}_F+\mathbf{K}_B,
\end{equation}
where the three terms on the right-hand side are stiffness from bar stretching, crease folding, and facet bending, respectively. For example, $ \mathbf{K}_S $ represents the in-plane stretching and shearing stiffness of the bar elements. Taking the bar element connecting pin-joints 2 and 2' as an example (Figure \ref{fig:Stability}(a)), one can define the bar stretching energy as
\begin{equation}
U_S^{22'}=\int_{0}^{L^{22'}} WA dX,
\end{equation}
where $ A $ is the bar element's cross-section area,  $ L^{22'} $  is the current bar length, and  $ W $ is the energy density, which is a function of the one dimensional Green-Lagrange strain $ E_{xx} = \textbf{B}_1\textbf{u}^{22'}+0.5\textbf{u}^{(22')T}\textbf{B}_2\textbf{u}^{22'} $. Here, the displacement vector of the bar element $ \textbf{u}^{22'}=\left[\textbf{d}_2^T \;\; \textbf{d}_{2'}^T\right]^T $, $ \textbf{B}_1=\left[-\textbf{e}_1  \;\; \textbf{e}_1\right]/L^{22'} $, $ \textbf{B}_2=\left[\textbf{I}_{3\times3} \;\; -\textbf{I}_{3\times3}; -\textbf{I}_{3\times3} \;\; \textbf{I}_{3\times3}\right]/(L^{22'})^2 $. $ \textbf{e}_1=\left[1,0,0\right] $, and $ \textbf{I}_{3\times3} $ is the identity matrix of size 3 by 3. The tangent stiffness matrix components corresponding to this bar element is

\vspace{-0.15in}
\begin{equation}\label{eq:KS}
\begin{split}
   \mathbf{K}_S^{22'} &=\frac{\partial^2 U_S^{22'}}{\partial \mathbf{u}^2} \\ &=k_s^{22'}L^{22'}\left(\mathbf{B}_1^T+\mathbf{B}_2 \mathbf{u}^{22'}\right)\left(\mathbf{B}_1^T+\mathbf{B}_2 \mathbf{u}^{22'}\right)^T+f^{22'}L^{22'}\mathbf{B}_2,
\end{split}
\end{equation}
\vspace{-0.2in}

\noindent where $ k_S^{22'} $ is the axial rigidity of the bar element, and $ f^{22'} $ is the resultant longitudinal force. It is worth noting that this stiffness matrix involves both the linear term and nonlinear terms related to geometry and initial displacement \cite{Liu2017e}. One can then apply similar formulations to all bar elements and assemble the global bar stiffness matrix.

The rotational hinges with prescribed torsional spring stiffness coefficients are applied to the bar elements corresponding to the folding creases and bending facets to approximate their deformation \cite{Liu2017e}.  As shown in Figure \ref{fig:Stability}(a), these torsional spring elements involve four vertices (nodes), five bars elements, and one dihedral angle between the two triangles defined by these bar elements.

Taking the rotational spring element corresponding to crease 1'-2' as an example (Figure \ref{fig:Stability}(a)). The crease folding energy is a function of the dihedral angle $ \varphi $
\begin{equation}
U_F^{1'2'}=\psi_{F}^{1'2'}(\varphi),
\end{equation}
where $ \psi_{F}^{1'2'} $ is the energy function. The dihedral angle between the two adjacent triangles (1'-2'-p and 1'-2'-q) can be calculated as $ \varphi=\eta \cos^{-1}\left(\frac{\mathbf{m} \cdot \mathbf{n}}{\left\| \mathbf{m} \right\| \left\| \mathbf{n} \right\|}\right) $, where the surface normal vectors $ \mathbf{m}=\mathbf{r}_{q1'} \times \mathbf{r}_{q2'} $, $ \mathbf{n}=\mathbf{r}_{p1'} \times \mathbf{r}_{p2'} $. The repeated indices do not imply summation in this paper. $ \eta $ is a sign indicator in that
\begin{equation}
\eta=\left\{
\begin{array}{rcl}
sgn(\mathbf{m} \cdot \mathbf{r}_{p2'}) & & {\mathbf{m} \cdot \mathbf{r}_{p2'} \neq 0}\\
1 & & {\mathbf{m} \cdot \mathbf{r}_{p2'} = 0}\\
\end{array} \right. .
\end{equation}

Because of the nonlinear geometric correlations among the dihedral angle ($ \varphi $) and the nodal displacement vector ($ \mathbf{u} $), the effective tangent stiffness is highly nonlinear even though its constituent creases are assumed to be linearly elastic in torsion \cite{Fang2017}. The tangent stiffness matrix component corresponding to the crease folding is

\begin{equation}
\label{eq:KF}
\mathbf{K}_F^{1'2'}=\frac{\partial^2 U_F^{1'2'}}{\partial \mathbf{u}^2}= k_f L^{1'2'} \frac{d\varphi}{d\mathbf{u}}\otimes \frac{d\varphi}{d\mathbf{u}} + M_f \frac{d^2\varphi}{d\mathbf{u}^2},
\end{equation}

\noindent where $ \otimes $ denotes the tensor product, $ L^{1'2'} $ is the length of the crease 1'-2', $ k_f $ is the torsional spring stiffness per unit length of the folding hinge, $ M_f $ is the rotational resistance moment, and $ \mathbf{u} $ is the nodal displacement vector of the related pin-joints at the current configuration.

The same formulation applies to the torsional spring elements corresponding to facet bending hinges because they have the same kinematic structure as a folded crease.  For example, in Figure \ref{fig:Stability}(a), the bending hinge with prescribed torsional spring stiffness $ k_{b} $ is assigned along the semi-diagonal of the quadrilateral facet (i.e., the bending line 1'-q).  The corresponding facet bending energy and tangent stiffness matrix are
\begin{align}
U_B^{1'q} & =\psi_{B}^{1'q}(\varphi) \\
\mathbf{K}_B^{1'q} & = k_{b}L^{1'q} \frac{d\varphi}{d\mathbf{u}}\otimes \frac{d\varphi}{d\mathbf{u}} + M_{b}\frac{d^2\varphi}{d\mathbf{u}^2}
\label{eq:KB}
\end{align}






\bibliographystyle{elsarticle-num-names}
\bibliography{Asymmetry_of_multi-stability.bib}

\begin{thebibliography}{50}
\expandafter\ifx\csname natexlab\endcsname\relax\def\natexlab#1{#1}\fi
\providecommand{\url}[1]{\texttt{#1}}
\providecommand{\href}[2]{#2}
\providecommand{\path}[1]{#1}
\providecommand{\DOIprefix}{doi:}
\providecommand{\ArXivprefix}{arXiv:}
\providecommand{\URLprefix}{URL: }
\providecommand{\Pubmedprefix}{pmid:}
\providecommand{\doi}[1]{\href{http://dx.doi.org/#1}{\path{#1}}}
\providecommand{\Pubmed}[1]{\href{pmid:#1}{\path{#1}}}
\providecommand{\bibinfo}[2]{#2}
\ifx\xfnm\relax \def\xfnm[#1]{\unskip,\space#1}\fi
\bibitem[{Iniguez-Rabago et~al.(2019)Iniguez-Rabago, Li, and
  Overvelde}]{Iniguez-Rabago2019}
\bibinfo{author}{A.~Iniguez-Rabago}, \bibinfo{author}{Y.~Li},
  \bibinfo{author}{J.~T. Overvelde},
\newblock \bibinfo{title}{{Exploring multistability in prismatic metamaterials
  through local actuation}},
\newblock \bibinfo{journal}{Nature Communications} \bibinfo{volume}{10}
  (\bibinfo{year}{2019}) \bibinfo{pages}{1--10}.
  \DOIprefix\doi{10.1038/s41467-019-13319-7}.
\bibitem[{Nadkarni et~al.(2016)Nadkarni, Arrieta, Chong, Kochmann, and
  Daraio}]{Nadkarni2016}
\bibinfo{author}{N.~Nadkarni}, \bibinfo{author}{A.~F. Arrieta},
  \bibinfo{author}{C.~Chong}, \bibinfo{author}{D.~M. Kochmann},
  \bibinfo{author}{C.~Daraio},
\newblock \bibinfo{title}{{Unidirectional Transition Waves in Bistable
  Lattices}},
\newblock \bibinfo{journal}{Physical Review Letters} \bibinfo{volume}{116}
  (\bibinfo{year}{2016}) \bibinfo{pages}{244501}.
  \DOIprefix\doi{10.1103/PhysRevLett.116.244501}.
\bibitem[{Raney et~al.(2016)Raney, Nadkarni, Daraio, Kochmann, Lewis, and
  Bertoldi}]{Raney2016a}
\bibinfo{author}{J.~R. Raney}, \bibinfo{author}{N.~Nadkarni},
  \bibinfo{author}{C.~Daraio}, \bibinfo{author}{D.~M. Kochmann},
  \bibinfo{author}{J.~A. Lewis}, \bibinfo{author}{K.~Bertoldi},
\newblock \bibinfo{title}{{Stable propagation of mechanical signals in soft
  media using stored elastic energy}},
\newblock \bibinfo{journal}{Proceedings of the National Academy of Sciences of
  the United States of America} \bibinfo{volume}{113} (\bibinfo{year}{2016})
  \bibinfo{pages}{9722--9727}. \DOIprefix\doi{10.1073/pnas.1604838113}.
\bibitem[{Daqaq et~al.(2014)Daqaq, Masana, Erturk, and Quinn}]{Daqaq2014}
\bibinfo{author}{M.~F. Daqaq}, \bibinfo{author}{R.~Masana},
  \bibinfo{author}{A.~Erturk}, \bibinfo{author}{D.~D. Quinn},
\newblock \bibinfo{title}{{On the role of nonlinearities in vibratory energy
  harvesting: A critical review and discussion}},
\newblock \bibinfo{journal}{Applied Mechanics Reviews} \bibinfo{volume}{66}
  (\bibinfo{year}{2014}). \DOIprefix\doi{10.1115/1.4026278}.
\bibitem[{Harne and Wang(2013)}]{Harne2013}
\bibinfo{author}{R.~L. Harne}, \bibinfo{author}{K.~W. Wang},
\newblock \bibinfo{title}{{A review of the recent research on vibration energy
  harvesting via bistable systems}},
\newblock \bibinfo{journal}{Smart Materials and Structures}
  \bibinfo{volume}{22} (\bibinfo{year}{2013}) \bibinfo{pages}{023001}.
  \DOIprefix\doi{10.1088/0964-1726/22/2/023001}.
\bibitem[{Pellegrini et~al.(2013)Pellegrini, Tolou, Schenk, and
  Herder}]{Pellegrini2013}
\bibinfo{author}{S.~P. Pellegrini}, \bibinfo{author}{N.~Tolou},
  \bibinfo{author}{M.~Schenk}, \bibinfo{author}{J.~L. Herder},
\newblock \bibinfo{title}{{Bistable vibration energy harvesters: A review}},
\newblock \bibinfo{journal}{Journal of Intelligent Material Systems and
  Structures} \bibinfo{volume}{24} (\bibinfo{year}{2013})
  \bibinfo{pages}{1303--1312}. \DOIprefix\doi{10.1177/1045389X12444940}.
\bibitem[{Daynes and Weaver(2013)}]{Daynes2013a}
\bibinfo{author}{S.~Daynes}, \bibinfo{author}{P.~M. Weaver},
\newblock \bibinfo{title}{{Review of shape-morphing automobile structures:
  Concepts and outlook}},
\newblock \bibinfo{journal}{Proceedings of the Institution of Mechanical
  Engineers, Part D: Journal of Automobile Engineering} \bibinfo{volume}{227}
  (\bibinfo{year}{2013}) \bibinfo{pages}{1603--1622}.
  \DOIprefix\doi{10.1177/0954407013496557}.
\bibitem[{{Xavier Lachenal, Stephen Daynes}(2013)}]{Lachenal2013}
\bibinfo{author}{P.~M.~W. {Xavier Lachenal, Stephen Daynes}},
\newblock \bibinfo{title}{{Review of morphing concepts and materials for wind
  turbine blade applications}},
\newblock \bibinfo{journal}{Wind Energy}  (\bibinfo{year}{2013})
  \bibinfo{pages}{1--20}. \DOIprefix\doi{10.1002/we}.
\bibitem[{Sun et~al.(2016)Sun, Guan, Liu, and Leng}]{Sun2016b}
\bibinfo{author}{J.~Sun}, \bibinfo{author}{Q.~Guan}, \bibinfo{author}{Y.~Liu},
  \bibinfo{author}{J.~Leng},
\newblock \bibinfo{title}{{Morphing aircraft based on smart materials and
  structures: A state-of-the-art review}},
\newblock \bibinfo{journal}{Journal of Intelligent Material Systems and
  Structures} \bibinfo{volume}{27} (\bibinfo{year}{2016})
  \bibinfo{pages}{2289--2312}. \DOIprefix\doi{10.1177/1045389X16629569}.
\bibitem[{Florijn et~al.(2014)Florijn, Coulais, and van Hecke}]{Florijn2014}
\bibinfo{author}{B.~Florijn}, \bibinfo{author}{C.~Coulais},
  \bibinfo{author}{M.~van Hecke},
\newblock \bibinfo{title}{{Programmable Mechanical Metamaterials}},
\newblock \bibinfo{journal}{Physical Review Letters} \bibinfo{volume}{113}
  (\bibinfo{year}{2014}) \bibinfo{pages}{175503}.
  \DOIprefix\doi{10.1103/PhysRevLett.113.175503}.
  \href{http://arxiv.org/abs/1407.4273}{{\tt arXiv:1407.4273}}.
\bibitem[{Kidambi et~al.(2016)Kidambi, Harne, and Wang}]{Kidambi2016}
\bibinfo{author}{N.~Kidambi}, \bibinfo{author}{R.~L. Harne},
  \bibinfo{author}{K.~W. Wang},
\newblock \bibinfo{title}{{Adaptation of Energy Dissipation in a Mechanical
  Metastable Module Excited Near Resonance}},
\newblock \bibinfo{journal}{Journal of Vibration and Acoustics}
  \bibinfo{volume}{138} (\bibinfo{year}{2016}) \bibinfo{pages}{1--9}.
  \DOIprefix\doi{10.1115/1.4031411}.
\bibitem[{Wu et~al.(2016)Wu, Harne, and Wang}]{Wu2016a}
\bibinfo{author}{Z.~Wu}, \bibinfo{author}{R.~L. Harne}, \bibinfo{author}{K.~W.
  Wang},
\newblock \bibinfo{title}{{Exploring a modular adaptive metastructure concept
  inspired by muscle's cross-bridge}},
\newblock \bibinfo{journal}{Journal of Intelligent Material Systems and
  Structures} \bibinfo{volume}{27} (\bibinfo{year}{2016})
  \bibinfo{pages}{1189--1202}. \DOIprefix\doi{10.1177/1045389X15586451}.
\bibitem[{Harne et~al.(2016)Harne, Wu, and Wang}]{Harne2016}
\bibinfo{author}{R.~L. Harne}, \bibinfo{author}{Z.~Wu}, \bibinfo{author}{K.~W.
  Wang},
\newblock \bibinfo{title}{{Designing and Harnessing the Metastable States of a
  Modular Metastructure for Programmable Mechanical Properties Adaptation}},
\newblock \bibinfo{journal}{Journal of Mechanical Design, Transactions of the
  ASME} \bibinfo{volume}{138} (\bibinfo{year}{2016}) \bibinfo{pages}{1--9}.
  \DOIprefix\doi{10.1115/1.4032093}.
\bibitem[{Grima et~al.(2013)Grima, Caruana-Gauci, Dudek, Wojciechowski, and
  Gatt}]{Grima2013}
\bibinfo{author}{J.~N. Grima}, \bibinfo{author}{R.~Caruana-Gauci},
  \bibinfo{author}{M.~R. Dudek}, \bibinfo{author}{K.~W. Wojciechowski},
  \bibinfo{author}{R.~Gatt},
\newblock \bibinfo{title}{{Smart metamaterials with tunable auxetic and other
  properties}},
\newblock \bibinfo{journal}{Smart Materials and Structures}
  \bibinfo{volume}{22} (\bibinfo{year}{2013}).
  \DOIprefix\doi{10.1088/0964-1726/22/8/084016}.
\bibitem[{Shan et~al.(2015)Shan, Kang, Raney, Wang, Fang, Candido, Lewis, and
  Bertoldi}]{Shan2015}
\bibinfo{author}{S.~Shan}, \bibinfo{author}{S.~H. Kang}, \bibinfo{author}{J.~R.
  Raney}, \bibinfo{author}{P.~Wang}, \bibinfo{author}{L.~Fang},
  \bibinfo{author}{F.~Candido}, \bibinfo{author}{J.~A. Lewis},
  \bibinfo{author}{K.~Bertoldi},
\newblock \bibinfo{title}{{Multistable Architected Materials for Trapping
  Elastic Strain Energy}},
\newblock \bibinfo{journal}{Advanced Materials} \bibinfo{volume}{27}
  (\bibinfo{year}{2015}) \bibinfo{pages}{4296--4301}. \URLprefix
  \url{http://doi.wiley.com/10.1002/adma.201501708}.
  \DOIprefix\doi{10.1002/adma.201501708}.
\bibitem[{Frenzel et~al.(2016)Frenzel, Findeisen, Kadic, Gumbsch, and
  Wegener}]{Frenzel2016}
\bibinfo{author}{T.~Frenzel}, \bibinfo{author}{C.~Findeisen},
  \bibinfo{author}{M.~Kadic}, \bibinfo{author}{P.~Gumbsch},
  \bibinfo{author}{M.~Wegener},
\newblock \bibinfo{title}{{Tailored Buckling Microlattices as Reusable
  Light-Weight Shock Absorbers}},
\newblock \bibinfo{journal}{Advanced Materials} \bibinfo{volume}{28}
  (\bibinfo{year}{2016}) \bibinfo{pages}{5865--5870}.
  \DOIprefix\doi{10.1002/adma.201600610}.
\bibitem[{Harne and Wang(2014)}]{Harne2014a}
\bibinfo{author}{R.~L. Harne}, \bibinfo{author}{K.~W. Wang},
\newblock \bibinfo{title}{{A bifurcation-based coupled linear-bistable system
  for microscale mass sensing}},
\newblock \bibinfo{journal}{Journal of Sound and Vibration}
  \bibinfo{volume}{333} (\bibinfo{year}{2014}) \bibinfo{pages}{2241--2252}.
  \URLprefix \url{http://dx.doi.org/10.1016/j.jsv.2013.12.017}.
  \DOIprefix\doi{10.1016/j.jsv.2013.12.017}.
\bibitem[{Harne and Wang(2015)}]{Harne2015}
\bibinfo{author}{R.~L. Harne}, \bibinfo{author}{K.~W. Wang},
\newblock \bibinfo{title}{{Passive measurement of progressive mass change via
  bifurcation sensing with a multistable micromechanical system}},
\newblock \bibinfo{journal}{Journal of Intelligent Material Systems and
  Structures} \bibinfo{volume}{26} (\bibinfo{year}{2015})
  \bibinfo{pages}{1622--1632}. \DOIprefix\doi{10.1177/1045389X14546781}.
\bibitem[{Kim et~al.(2014)Kim, Koh, Lee, Ryu, Cho, and Cho}]{Kim2014a}
\bibinfo{author}{S.-W.~W. Kim}, \bibinfo{author}{J.-S.~S. Koh},
  \bibinfo{author}{J.-G.~G. Lee}, \bibinfo{author}{J.~Ryu},
  \bibinfo{author}{M.~Cho}, \bibinfo{author}{K.-J.~J. Cho},
\newblock \bibinfo{title}{{Flytrap-inspired robot using structurally integrated
  actuation based on bistability and a developable surface}},
\newblock \bibinfo{journal}{Bioinspiration and Biomimetics} \bibinfo{volume}{9}
  (\bibinfo{year}{2014}) \bibinfo{pages}{036004}.
  \DOIprefix\doi{10.1088/1748-3182/9/3/036004}.
\bibitem[{Chen et~al.(2018)Chen, Bilal, Shea, and Daraio}]{Chen2018a}
\bibinfo{author}{T.~Chen}, \bibinfo{author}{O.~R. Bilal},
  \bibinfo{author}{K.~Shea}, \bibinfo{author}{C.~Daraio},
\newblock \bibinfo{title}{{Harnessing bistability for directional propulsion of
  soft, untethered robots}},
\newblock \bibinfo{journal}{Proceedings of the National Academy of Sciences of
  the United States of America} \bibinfo{volume}{115} (\bibinfo{year}{2018})
  \bibinfo{pages}{5698--5702}. \DOIprefix\doi{10.1073/pnas.1800386115}.
\bibitem[{Bhovad et~al.(2019)Bhovad, Kaufmann, and Li}]{Bhovad2019a}
\bibinfo{author}{P.~Bhovad}, \bibinfo{author}{J.~Kaufmann},
  \bibinfo{author}{S.~Li},
\newblock \bibinfo{title}{{Peristaltic locomotion without digital controllers:
  Exploiting multi-stability in origami to coordinate robotic motion}},
\newblock \bibinfo{journal}{Extreme Mechanics Letters} \bibinfo{volume}{32}
  (\bibinfo{year}{2019}) \bibinfo{pages}{100552}. \URLprefix
  \url{https://doi.org/10.1016/j.eml.2019.100552}.
  \DOIprefix\doi{10.1016/j.eml.2019.100552}.
\bibitem[{Preston et~al.(2019)Preston, Jiang, Sanchez, Rothemund, Rawson,
  Nemitz, Lee, Suo, Walsh, and Whitesides}]{Preston2019}
\bibinfo{author}{D.~J. Preston}, \bibinfo{author}{H.~J. Jiang},
  \bibinfo{author}{V.~Sanchez}, \bibinfo{author}{P.~Rothemund},
  \bibinfo{author}{J.~Rawson}, \bibinfo{author}{M.~P. Nemitz},
  \bibinfo{author}{W.~K. Lee}, \bibinfo{author}{Z.~Suo}, \bibinfo{author}{C.~J.
  Walsh}, \bibinfo{author}{G.~M. Whitesides},
\newblock \bibinfo{title}{{A soft ring oscillator}},
\newblock \bibinfo{journal}{Science Robotics} \bibinfo{volume}{4}
  (\bibinfo{year}{2019}) \bibinfo{pages}{1--10}.
  \DOIprefix\doi{10.1126/scirobotics.aaw5496}.
\bibitem[{Sengupta and Li(2018)}]{Sengupta2018a}
\bibinfo{author}{S.~Sengupta}, \bibinfo{author}{S.~Li},
\newblock \bibinfo{title}{{Harnessing the anisotropic multistability of
  stacked-origami mechanical metamaterials for effective modulus programming}},
\newblock \bibinfo{journal}{Journal of Intelligent Material Systems and
  Structures} \bibinfo{volume}{29} (\bibinfo{year}{2018})
  \bibinfo{pages}{2933--2945}. \DOIprefix\doi{10.1177/1045389X18781040}.
\bibitem[{Lele et~al.(2019)Lele, Deshpande, Myers, and Li}]{Lele2019a}
\bibinfo{author}{A.~Lele}, \bibinfo{author}{V.~Deshpande},
  \bibinfo{author}{O.~Myers}, \bibinfo{author}{S.~Li},
\newblock \bibinfo{title}{{Snap-through and stiffness adaptation of a
  multi-stable Kirigami composite module}},
\newblock \bibinfo{journal}{Composites Science and Technology}
  \bibinfo{volume}{182} (\bibinfo{year}{2019}) \bibinfo{pages}{107750}.
  \URLprefix \url{https://doi.org/10.1016/j.compscitech.2019.107750}.
  \DOIprefix\doi{10.1016/j.compscitech.2019.107750}.
\bibitem[{Peraza-Hernandez et~al.(2014)Peraza-Hernandez, Hartl, Malak, and
  Lagoudas}]{Peraza-Hernandez2014}
\bibinfo{author}{E.~A. Peraza-Hernandez}, \bibinfo{author}{D.~J. Hartl},
  \bibinfo{author}{R.~J. Malak}, \bibinfo{author}{D.~C. Lagoudas},
\newblock \bibinfo{title}{{Origami-inspired active structures: A synthesis and
  review}},
\newblock \bibinfo{journal}{Smart Materials and Structures}
  \bibinfo{volume}{23} (\bibinfo{year}{2014}).
  \DOIprefix\doi{10.1088/0964-1726/23/9/094001}.
\bibitem[{Johnson et~al.(2017)Johnson, Chen, Hovet, Xu, Wood, Ren, Tokuda, and
  Tse}]{Johnson2017}
\bibinfo{author}{M.~Johnson}, \bibinfo{author}{Y.~Chen},
  \bibinfo{author}{S.~Hovet}, \bibinfo{author}{S.~Xu},
  \bibinfo{author}{B.~Wood}, \bibinfo{author}{H.~Ren},
  \bibinfo{author}{J.~Tokuda}, \bibinfo{author}{Z.~T.~H. Tse},
\newblock \bibinfo{title}{{Fabricating biomedical origami: a state-of-the-art
  review}},
\newblock \bibinfo{journal}{International Journal of Computer Assisted
  Radiology and Surgery} \bibinfo{volume}{12} (\bibinfo{year}{2017})
  \bibinfo{pages}{2023--2032}. \DOIprefix\doi{10.1007/s11548-017-1545-1}.
\bibitem[{Rus and Tolley(2018)}]{Rus2018}
\bibinfo{author}{D.~Rus}, \bibinfo{author}{M.~T. Tolley},
\newblock \bibinfo{title}{{Design, fabrication and control of origami robots}},
\newblock \bibinfo{journal}{Nature Reviews Materials} \bibinfo{volume}{3}
  (\bibinfo{year}{2018}) \bibinfo{pages}{101--112}. \URLprefix
  \url{http://dx.doi.org/10.1038/s41578-018-0009-8}.
  \DOIprefix\doi{10.1038/s41578-018-0009-8}.
\bibitem[{Waitukaitis et~al.(2015)Waitukaitis, Menaut, Chen, and van
  Hecke}]{Waitukaitis2015}
\bibinfo{author}{S.~Waitukaitis}, \bibinfo{author}{R.~Menaut},
  \bibinfo{author}{B.~G.-g. Chen}, \bibinfo{author}{M.~van Hecke},
\newblock \bibinfo{title}{{Origami Multistability: From Single Vertices to
  Metasheets}},
\newblock \bibinfo{journal}{Physical Review Letters} \bibinfo{volume}{114}
  (\bibinfo{year}{2015}) \bibinfo{pages}{055503}.
  \DOIprefix\doi{10.1103/PhysRevLett.114.055503}.
\bibitem[{Yasuda and Yang(2015)}]{Yasuda2015a}
\bibinfo{author}{H.~Yasuda}, \bibinfo{author}{J.~Yang},
\newblock \bibinfo{title}{{Reentrant Origami-Based Metamaterials with Negative
  Poisson's Ratio and Bistability}},
\newblock \bibinfo{journal}{Physical Review Letters} \bibinfo{volume}{114}
  (\bibinfo{year}{2015}) \bibinfo{pages}{185502}. \URLprefix
  \url{https://link.aps.org/doi/10.1103/PhysRevLett.114.185502}.
  \DOIprefix\doi{10.1103/PhysRevLett.114.185502}.
\bibitem[{Yasuda et~al.(2016)Yasuda, Chen, and Yang}]{Yasuda2016}
\bibinfo{author}{H.~Yasuda}, \bibinfo{author}{Z.~Chen},
  \bibinfo{author}{J.~Yang},
\newblock \bibinfo{title}{{Multitransformable Leaf-Out Origami With Bistable
  Behavior}},
\newblock \bibinfo{journal}{Journal of Mechanisms and Robotics}
  \bibinfo{volume}{8} (\bibinfo{year}{2016}) \bibinfo{pages}{031013}.
  \DOIprefix\doi{10.1115/1.4031809}.
\bibitem[{Sadeghi and Li(2020)}]{Sadeghi2020}
\bibinfo{author}{S.~Sadeghi}, \bibinfo{author}{S.~Li},
\newblock \bibinfo{title}{{Dynamic Folding of Origami By Exploiting Asymmetric
  Multi-Stability}},
\newblock \bibinfo{journal}{arXiv}  (\bibinfo{year}{2020})
  \bibinfo{pages}{1--24}. \href{http://arxiv.org/abs/2006.05968}{{\tt
  arXiv:2006.05968}}.
\bibitem[{Kaufmann et~al.(2021)Kaufmann, Bhovad, and Li}]{Kaufmann2021a}
\bibinfo{author}{J.~Kaufmann}, \bibinfo{author}{P.~Bhovad},
  \bibinfo{author}{S.~Li},
\newblock \bibinfo{title}{{Harnessing the Multistability of Kresling Origami
  for Reconfigurable Articulation in Soft Robotic Arms}},
\newblock \bibinfo{journal}{Soft Robotics} \bibinfo{volume}{00}
  (\bibinfo{year}{2021}) \bibinfo{pages}{1--12}.
  \DOIprefix\doi{10.1089/soro.2020.0075}.
  \href{http://arxiv.org/abs/2008.07421}{{\tt arXiv:2008.07421}}.
\bibitem[{Ma et~al.(2021)Ma, Zang, Feng, Chen, and You}]{Ma2021}
\bibinfo{author}{J.~Ma}, \bibinfo{author}{S.~Zang}, \bibinfo{author}{H.~Feng},
  \bibinfo{author}{Y.~Chen}, \bibinfo{author}{Z.~You},
\newblock \bibinfo{title}{{Theoretical characterization of a non-rigid-foldable
  square-twist origami for property programmability}},
\newblock \bibinfo{journal}{International Journal of Mechanical Sciences}
  \bibinfo{volume}{189} (\bibinfo{year}{2021}) \bibinfo{pages}{105981}.
  \DOIprefix\doi{10.1016/j.ijmecsci.2020.105981}.
\bibitem[{Kamrava et~al.(2019)Kamrava, Ghosh, Wang, and Vaziri}]{Kamrava2019}
\bibinfo{author}{S.~Kamrava}, \bibinfo{author}{R.~Ghosh},
  \bibinfo{author}{Z.~Wang}, \bibinfo{author}{A.~Vaziri},
\newblock \bibinfo{title}{{Origami-Inspired Cellular Metamaterial With
  Anisotropic Multi-Stability}},
\newblock \bibinfo{journal}{Advanced Engineering Materials}
  \bibinfo{volume}{21} (\bibinfo{year}{2019}) \bibinfo{pages}{1800895}.
  \DOIprefix\doi{10.1002/adem.201800895}.
\bibitem[{Melancon et~al.(2021)Melancon, Gorissen, Garc{\'{i}}a-Mora, Hoberman,
  and Bertoldi}]{Melancon2021}
\bibinfo{author}{D.~Melancon}, \bibinfo{author}{B.~Gorissen},
  \bibinfo{author}{C.~J. Garc{\'{i}}a-Mora}, \bibinfo{author}{C.~Hoberman},
  \bibinfo{author}{K.~Bertoldi},
\newblock \bibinfo{title}{{Multistable inflatable origami structures at the
  metre scale}},
\newblock \bibinfo{journal}{Nature} \bibinfo{volume}{592}
  (\bibinfo{year}{2021}) \bibinfo{pages}{545--550}. \URLprefix
  \url{http://dx.doi.org/10.1038/s41586-021-03407-4}.
  \DOIprefix\doi{10.1038/s41586-021-03407-4}.
\bibitem[{Silverberg et~al.(2015)Silverberg, Na, Evans, Liu, Hull, Santangelo,
  Lang, Hayward, and Cohen}]{Silverberg2015}
\bibinfo{author}{J.~L. Silverberg}, \bibinfo{author}{J.-H. Na},
  \bibinfo{author}{A.~A. Evans}, \bibinfo{author}{B.~Liu},
  \bibinfo{author}{T.~C. Hull}, \bibinfo{author}{C.~D. Santangelo},
  \bibinfo{author}{R.~J. Lang}, \bibinfo{author}{R.~C. Hayward},
  \bibinfo{author}{I.~Cohen},
\newblock \bibinfo{title}{{Origami structures with a critical transition to
  bistability arising from hidden degrees of freedom}},
\newblock \bibinfo{journal}{Nature Materials} \bibinfo{volume}{14}
  (\bibinfo{year}{2015}) \bibinfo{pages}{389--393}. \URLprefix
  \url{http://goo.gl/kk3lXk http://www.nature.com/articles/nmat4232}.
  \DOIprefix\doi{10.1038/nmat4232}.
\bibitem[{Pagano et~al.(2017)Pagano, Yan, Chien, Wissa, and
  Tawfick}]{Pagano2017}
\bibinfo{author}{A.~Pagano}, \bibinfo{author}{T.~Yan},
  \bibinfo{author}{B.~Chien}, \bibinfo{author}{A.~Wissa},
  \bibinfo{author}{S.~Tawfick},
\newblock \bibinfo{title}{{A crawling robot driven by multi-stable origami}},
\newblock \bibinfo{journal}{Smart Materials and Structures}
  \bibinfo{volume}{26} (\bibinfo{year}{2017}).
  \DOIprefix\doi{10.1088/1361-665X/aa721e}.
\bibitem[{Liu and Paulino(2018)}]{Liu2018}
\bibinfo{author}{K.~Liu}, \bibinfo{author}{G.~H. Paulino},
\newblock \bibinfo{title}{{Highly efficient nonlinear structural analysis of
  origami assemblages using the MERLIN2 software}},
\newblock \bibinfo{journal}{The proceedings from the seventh meeting of
  Origami, Science, Mathematics and Education}  (\bibinfo{year}{2018}).
\bibitem[{Schenk and Guest(2013)}]{Schenk2013a}
\bibinfo{author}{M.~Schenk}, \bibinfo{author}{S.~D. Guest},
\newblock \bibinfo{title}{{Geometry of Miura-folded metamaterials}},
\newblock \bibinfo{journal}{Proceedings of the National Academy of Sciences}
  \bibinfo{volume}{110} (\bibinfo{year}{2013}) \bibinfo{pages}{3276--3281}.
  \DOIprefix\doi{10.1073/pnas.1217998110}.
\bibitem[{Fang et~al.(2016)Fang, Li, Ji, and Wang}]{Fang2016b}
\bibinfo{author}{H.~Fang}, \bibinfo{author}{S.~Li}, \bibinfo{author}{H.~Ji},
  \bibinfo{author}{K.~W. Wang},
\newblock \bibinfo{title}{{Uncovering the deformation mechanisms of origami
  metamaterials by introducing generic degree-four vertices}},
\newblock \bibinfo{journal}{Physical Review E} \bibinfo{volume}{94}
  (\bibinfo{year}{2016}) \bibinfo{pages}{1--11}.
  \DOIprefix\doi{10.1103/PhysRevE.94.043002}.
\bibitem[{Li and Wang(2015)}]{Li2015b}
\bibinfo{author}{S.~Li}, \bibinfo{author}{K.~W. Wang},
\newblock \bibinfo{title}{{Fluidic origami: a plant-inspired adaptive structure
  with shape morphing and stiffness tuning}},
\newblock \bibinfo{journal}{Smart Materials and Structures}
  \bibinfo{volume}{24} (\bibinfo{year}{2015}) \bibinfo{pages}{105031}.
  \DOIprefix\doi{10.1088/0964-1726/24/10/105031}.
\bibitem[{Sane et~al.(2018)Sane, Bhovad, and Li}]{Sane2018a}
\bibinfo{author}{H.~Sane}, \bibinfo{author}{P.~Bhovad},
  \bibinfo{author}{S.~Li},
\newblock \bibinfo{title}{{Actuation performance of fluidic origami cellular
  structure: A holistic investigation}},
\newblock \bibinfo{journal}{Smart Materials and Structures}
  \bibinfo{volume}{27} (\bibinfo{year}{2018}).
  \DOIprefix\doi{10.1088/1361-665X/aadfac}.
\bibitem[{Fang et~al.(2018)Fang, Chu, Xia, and Wang}]{Fang2018}
\bibinfo{author}{H.~Fang}, \bibinfo{author}{S.~C.~A. Chu},
  \bibinfo{author}{Y.~Xia}, \bibinfo{author}{K.~W. Wang},
\newblock \bibinfo{title}{{Programmable Self-Locking Origami Mechanical
  Metamaterials}},
\newblock \bibinfo{journal}{Advanced Materials} \bibinfo{volume}{30}
  (\bibinfo{year}{2018}) \bibinfo{pages}{1--9}.
  \DOIprefix\doi{10.1002/adma.201706311}.
\bibitem[{Li and Wang(2015)}]{Li2015c}
\bibinfo{author}{S.~Li}, \bibinfo{author}{K.~W. Wang},
\newblock \bibinfo{title}{{Fluidic origami with embedded pressure dependent
  multi-stability: A plant inspired innovation}},
\newblock \bibinfo{journal}{Journal of the Royal Society Interface}
  \bibinfo{volume}{12} (\bibinfo{year}{2015}).
  \DOIprefix\doi{10.1098/rsif.2015.0639}.
\bibitem[{Fang et~al.(2017)Fang, Wang, and Li}]{Fang2017}
\bibinfo{author}{H.~Fang}, \bibinfo{author}{K.~Wang}, \bibinfo{author}{S.~Li},
\newblock \bibinfo{title}{{Asymmetric energy barrier and mechanical diode
  effect from folding multi-stable stacked-origami}},
\newblock \bibinfo{journal}{Extreme Mechanics Letters} \bibinfo{volume}{17}
  (\bibinfo{year}{2017}) \bibinfo{pages}{7--15}.
  \DOIprefix\doi{10.1016/j.eml.2017.09.008}.
\bibitem[{Gillman et~al.(2018)Gillman, Fuchi, and Buskohl}]{Gillman2018e}
\bibinfo{author}{A.~Gillman}, \bibinfo{author}{K.~Fuchi},
  \bibinfo{author}{P.~Buskohl},
\newblock \bibinfo{title}{{Truss-based nonlinear mechanical analysis for
  origami structures exhibiting bifurcation and limit point instabilities}},
\newblock \bibinfo{journal}{International Journal of Solids and Structures}
  \bibinfo{volume}{147} (\bibinfo{year}{2018}) \bibinfo{pages}{80--93}.
  \DOIprefix\doi{10.1016/j.ijsolstr.2018.05.011}.
\bibitem[{Liu et~al.(2019)Liu, Tachi, and Paulino}]{Liu2019a}
\bibinfo{author}{K.~Liu}, \bibinfo{author}{T.~Tachi}, \bibinfo{author}{G.~H.
  Paulino},
\newblock \bibinfo{title}{{Invariant and smooth limit of discrete geometry
  folded from bistable origami leading to multistable metasurfaces}},
\newblock \bibinfo{journal}{Nature Communications} \bibinfo{volume}{10}
  (\bibinfo{year}{2019}) \bibinfo{pages}{4238}.
  \DOIprefix\doi{10.1038/s41467-019-11935-x}.
\bibitem[{Filipov et~al.(2017)Filipov, Liu, Tachi, Schenk, and
  Paulino}]{Filipov2017}
\bibinfo{author}{E.~Filipov}, \bibinfo{author}{K.~Liu},
  \bibinfo{author}{T.~Tachi}, \bibinfo{author}{M.~Schenk},
  \bibinfo{author}{G.~Paulino},
\newblock \bibinfo{title}{{Bar and hinge models for scalable analysis of
  origami}},
\newblock \bibinfo{journal}{International Journal of Solids and Structures}
  \bibinfo{volume}{124} (\bibinfo{year}{2017}) \bibinfo{pages}{26--45}.
  \DOIprefix\doi{10.1016/j.ijsolstr.2017.05.028}.
\bibitem[{Liu and Paulino(2017)}]{Liu2017e}
\bibinfo{author}{K.~Liu}, \bibinfo{author}{G.~H. Paulino},
\newblock \bibinfo{title}{{Nonlinear mechanics of non-rigid origami: an
  efficient computational approach}},
\newblock \bibinfo{journal}{Proceedings of the Royal Society A: Mathematical,
  Physical and Engineering Sciences} \bibinfo{volume}{473}
  (\bibinfo{year}{2017}) \bibinfo{pages}{20170348}.
  \DOIprefix\doi{10.1098/rspa.2017.0348}.
\bibitem[{Baharisangari and Li(2019)}]{Baharisangari2019c}
\bibinfo{author}{N.~Baharisangari}, \bibinfo{author}{S.~Li},
\newblock \bibinfo{title}{{Exploiting the asymmetric energy barrier in
  multi-stable origami to enable mechanical diode behavior in compression}},
\newblock \bibinfo{journal}{Proceedings of the ASME Design Engineering
  Technical Conference} \bibinfo{volume}{5B-2019} (\bibinfo{year}{2019})
  \bibinfo{pages}{1--8}. \DOIprefix\doi{10.1115/DETC2019-97420}.

\end{thebibliography}







\end{document}